\documentstyle[aps,eqsecnum,epsfig,amsmath,amssymb]{revtex}

\def\bsy{\boldsymbol}
\def\eps{\varepsilon}
\def\phidag{{\bsy{\Phi}}^{\dag}}
\def\RE{\Bbb{R}}

\begin{document} 

\bibliographystyle{prsty} 

\draft

\title{Fronts, Domain Growth and Dynamical Scaling in a $\mathbf{d=1}$
non-Potential System} 

\author{ R. Gallego, M. San Miguel and R. Toral} 

\address{Instituto Mediterr\'aneo de Estudios Avanzados, IMEDEA 
\footnote{URL: http://www.imedea.uib.es/PhysDept}(CSIC-UIB) \\
Campus Universitat de les Illes Balears, E-07071 \\
Palma de Mallorca, Spain} 

\date{today} 

\maketitle

%%%%%%%%%%%%%%%%%%
\begin{abstract}
We present a study of dynamical scaling and front motion in a one dimensional
system that describes Rayleigh-B\'enard convection in a rotating cell. We  use
a model of three competing modes proposed by Busse and Heikes to which spatial
dependent terms have been added.  As long as the angular velocity is different
from zero, there is no known Lyapunov potential for the dynamics of the system.
As a consequence the system follows a non-relaxational dynamics and the
asymptotic state can not be associated with a final equilibrium state. When the
rotation angular velocity is greater than some critical value, the system
undergoes the K\"uppers-Lortz instability leading to a time dependent chaotic
dynamics and there is no coarsening beyond this instability. We have focused on
the transient dynamics below this instability, where the dynamics is still 
non-relaxational. In this regime the dynamics is governed by a non-relaxational
motion of fronts separating dynamically equivalent homogeneous states. We
classify the families of  fronts that occur in the dynamics, and calculate
their shape and velocity. We have found that a scaling description of the
coarsening process is still valid as in the  potential case. The growth law is
nearly logarithmic with time for short times and becomes linear after a
crossover, whose width is determined by the strength of the non-potential
terms.
\end{abstract} 
%%%%%%%%%%%%%%%%%%

\vskip 0.4cm

\section{Introduction}
\label{intro}

A noteworthy result in the study of non-equilibrium statistical mechanics is
the existence of dynamical scaling during the coarsening process in which a
system approaches equilibrium after undergoing a phase transition
\cite{GSMS,BrayReport}. Dynamical scaling reflects that domain growth is
self-similar with a single time dependent characteristic length. In the
simplest case of a relaxational dynamics for a scalar order parameter, which
models, for example, an order-disorder transition (model A in the taxonomy of
\cite{HH}),  domains of two equivalent phases grow locally from an unstable
state, and the approach to a final  equilibrium state is dominated by interface
motion. For spatial dimension  $d>1$, the mechanism for domain growth is
curvature driven interface motion. This leads to a characteristic length 
growing as $R\sim t^{1/2}$. This type of phenomena has been studied in a large
variety of systems which share the common feature that the final state of the
dynamics is a state of thermodynamic equilibrium which minimizes a free energy
or ``potential'' of the problem. Transient dynamics might also include
additional processes beyond pure relaxation in that potential
\cite{MontagnePhysica,SMTChile}, but a measure of relative stability between
stationary states is guaranteed by the existence of the potential. A more
genuine non-equilibrium dynamics occurs when such a potential does not exist. A
natural question, which we address in this paper, is the existence of dynamical
scaling in the approach to a final stationary state  which does not follow the
minimization of a potential,  while this transient  dynamics from an unstable
state involves the formation of spatial domains. The question of dynamical
scaling can also be addressed for Hamiltonian dynamics \cite{Rica}, which is
the extreme opposite situation to that of  dissipative relaxational dynamics in
a potential. A general non-equilibrium situation will, in general, have
contributions from both types of dynamics \cite{MontagnePhysica,SMTChile}.

The motion of an interface between two linearly stable solutions of a dynamical
system was long ago proposed as a measure of relative stability for a
non-potential system \cite{Kramer}, and the motion of interfaces, domain walls
or front solutions has been studied in a number of non-potential systems
\cite{Coullet1,Coullet2,Fauve,Meron}. It is known that a  domain wall between
two equivalent states with different broken symmetry can move in $d=1$ in
either direction due to non-potential dynamics \cite{Coullet1}. Likewise,
non-potential dynamics can stabilize front solutions which, in the potential
limit, would move from a globally stable into a metastable state \cite{Fauve}.
It is the purpose of this paper to study the consequences of interface motion
driven by non-potential dynamics on coarsening processes and dynamical scaling,
which have been by and large not considered.

A physical prototype situation in which the question posed above can be studied
is Rayleigh-B\'enard convection in a rotating cell \cite{AhlersEcke}, which in
many aspects is mathematically equivalent to a biological model of competition
between three species \cite{May,Frachebourg}.  Beyond a threshold value for the
parameter, $\delta$, measuring the strength of non-potential terms (related to
the rotation speed  in the case of Rayleigh--B\'enard convection) an
instability to a time  dependent dynamics occurs (K\"uppers-Lortz instability
\cite{Kuppers,Busse-Heikes}). Below this instability, but still taking the
system beyond the Rayleigh--B\'enard convective instability, locally ordered
domains, associated with different orientations of the convective rolls emerge.
The subsequent coarsening process seems to be stopped by the K\"uppers-Lortz 
instability in $d=2$ \cite{TuCross1,TuCross2}.  However, below the  instability
to a time dependent state, three preferred orientations exist and the motion of
interfaces separating them is subject to non-potential dynamics which will
affect domain growth.

With this motivation in mind, and as a first step towards the understanding of
the problem of domain growth and dynamical scaling in this type of systems, we
have considered a $d=1$ model for three competing real non-conserved order
parameters with the non-potential dynamics of \cite{May,Busse-Heikes} and 
spatial inhomogeneities determined by short range self-interactions.  We have
chosen this system because dynamical scaling in $d=1$ potential systems is a
special case for which well established results are available
\cite{BrayReport}: for the simplest case of a scalar non-conserved order
parameter with short range interactions, a scaling solution is known with a
logarithmic growth law for the typical domain size, $R \sim \log
t$~\cite{Rutenberg,Nagai}. This regime follows an early time regime of domain
formation with a growth law $R\sim t^{1/2}$~\cite{DePasquale}. The logarithmic
domain growth has its origin in the interactions between domain
walls~\cite{KawasakiOhta}. The velocity of domain wall motion in these
circumstances can be calculated by a perturbation analysis  \cite{Coullet3}. We
study how these results are modified by non-potential dynamics. We find that
dynamical scaling still holds, but with a crossover between two well defined
regimes characterized by a logarithmic and linear domain growth law
respectively. The two growth laws can be traced back to the two mechanisms that
determine domain wall motion. The first one is the interaction between domain
walls  as in the potential case. The second one is due to the fact that  the
non-potential dynamics causes that isolated individual fronts move with finite
velocity. In a multi-front configuration this provides and additional
coarsening mechanism in which  fronts moving in opposite directions annihilate
each other. The crossover time between the two dominant mechanisms described
depends on the strength of non-potential terms. When these become large enough,
the logarithmic regime is pushed to just the very early times. In this case
finite size effects become also important since very large domains emerge
rather fast.

The outline of the paper is as follows: in section \ref{Model} we introduce the
non-potential model and describe its homogeneous stationary solutions. In
section \ref{Front} we clasify non-homogeneous front solutions and compute
their velocity. We also analyze the interactions between two fronts including
the  non-potential effects. In section \ref{Growth} we discuss the issue of
dynamical scaling and present numerical simulations that show the validity of
the dynamical scaling description when non-potential terms are present.
Finally, in section \ref{theend} we end with some conclusions and an outlook.

%%%%%%%%%%%%%%%%%%%%%%%%%%%%
\section{Theoretical model}
\label{Model}
%%%%%%%%%%%%%%%%%%%%%%%%%%%%

We base our theoretical approach in a three mode model first proposed in the
context of fluid dynamics by Busse and Heikes~\cite{Busse-Heikes}. In this
model, the (real)  amplitudes of the three selected modes corresponding to
three different orientations of the convection rolls, $A_{1}$, $A_{2}$,
$A_{3}$, follow the evolution equations:
\begin{equation}\label{eq:modelo}
  \begin{aligned}
  \partial_{t} A_{1} &=  A_{1}\, (1-A_{1}^{2}-(\eta + \delta)\, A_{2}^{2}
  -(\eta-\delta)\, A_{3}^{2})	\\
  \partial_{t} A_{2} &=  A_{2}\, (1-A_{2}^{2}-(\eta + \delta)\, A_{3}^{2}
  -(\eta-\delta)\, A_{1}^{2})	\\
  \partial_{t} A_{3} &=  A_{3}\, (1-A_{3}^{2}-(\eta + \delta)\, A_{1}^{2}
  -(\eta-\delta)\, A_{2}^{2})
  \end{aligned}
\end{equation}
A similar set of equations was proposed earlier to model the behavior of three 
competing biological species \cite{May}. In this case, $A_{1}$, $A_{2}$ and
$A_{3}$ stand for the population number of each species. For the fluid case, 
$\delta$ is related to the rotation speed such that $\delta=0$ is the
non-rotating case; the parameter $\eta$ is related to other fluid parameters.
The analysis of \cite{Busse-Heikes} and \cite{May} shows that, for a certain 
range of the parameters $\eta$ and $\delta$ (see next section) there are no
homogeneous stable solutions and the dynamics tends asymptotically to a
sequence of alternations of the three modes as shown in experiments. An
unwanted feature of the previous model is that the alternation time is not
constant, but increases with time, contrary to experiments where an
approximately constant period is observed. Although Busse and Heikes proposed
that the addition of small noise could stabilize the period 
\cite{SMTChile,TSM}, an alternative explanation considered the addition of
spatially dependent terms to the previous equations. Although the symmetries
that must satisfy the amplitude equations would imply that the spatially
dependent terms should be of an specific form~\cite{Newell,Segel,GOS}, it has
been shown in \cite{TuCross1} that these can be further simplified without 
altering the essentials of the problem. We choose in this work the simplest
diffusive form for the spatial depending terms, namely:
\begin{equation}\label{eq:model}
  \begin{aligned}
  \partial_{t} A_{1} &= \partial_x^{2} A_{1} + A_{1}\, (1-A_{1}^{2}-
  (\eta + \delta)\, A_{2}^{2}-(\eta-\delta)\, A_{3}^{2})	\\
  \partial_{t} A_{2} &= \partial_x^{2} A_{2} + A_{2}\, (1-A_{2}^{2}-
  (\eta + \delta)\, A_{3}^{2}-(\eta-\delta)\, A_{1}^{2})	\\
  \partial_{t} A_{3} &= \partial_x^{2} A_{3} + A_{3}\, (1-A_{3}^{2}-
  (\eta + \delta)\, A_{1}^{2}-(\eta-\delta)\, A_{2}^{2})
  \end{aligned}
\end{equation}
which form the basis of our subsequent analysis. A similar set of equations for
the modulus square of the amplitudes is introduced in
reference~\cite{Frachebourg} for some particular values of the parameters
$\eta$, $\delta$. Notice that  the system is invariant under the following
transformations: \\ a) $x\rightarrow x+x_{0}$, $t\rightarrow t+t_{0}$
(spatio-temporal  translation symmetry)\\ b) $A_{1}\rightarrow A_{2},\
A_{2}\rightarrow A_{3},\ A_{3}\rightarrow A_{1}$ (cyclic permutation
symmetry)\\ c) $A_{i}\rightleftarrows A_{j},\ \delta\rightarrow -\delta$, where
$A_{i},\ A_{j}$ are any two different amplitudes.\\ The previous analysis shows
that $A_{1}$, $A_{2}$ and $A_{3}$ are ``equivalent" (from the dynamical point
of view) variables.

It is possible to split the dynamical equations into a potential and a
non-potential contributions:
\begin{equation}\label{eq:model2}
    \partial_{t} A_{i} = -\frac{\delta {\mathcal F}}{\delta A_{i}}-
    \delta\cdot f_{i},\qquad i=1,2,3 
\end{equation}
where the potential function $\mathcal F$ is given by:
\begin{equation}
  \begin{split}\label{eq:Lyap}
    {\mathcal F}[A_{1},A_{2},A_{3}] &= - \int dx \left\{ \frac{1}{2}
    \left[ (\partial_{x}A_{1})^{2}+(\partial_{x}A_{2})^{2}+
    (\partial_{x}A_{3})^{2}\right]+
    \frac{1}{2} (A_{1}^{2}+A_{2}^{2}+A_{3}^{2}) \right. 		\\[-.1cm]
    &\phantom{xxxx} \left. -\frac{1}{4} (A_{1}^{4}+A_{2}^{4}+A_{3}^{4})
    -\frac{1}{2} \, \eta \, (A_{1}^{2}A_{2}^{2}+A_{1}^{2}A_{3}^{2}+
    A_{2}^{2}A_{3}^{2}) \right\}
  \end{split}
\end{equation}
and the non-potential terms are:
\begin{equation}
\begin{aligned}
    f_{1} &=  A_{1}(A_{2}^{2}-A_{3}^{2})	 \\
    f_{2} &=  A_{2}(A_{1}^{2}-A_{3}^{2})	 \\
    f_{3} &=  A_{3}(A_{2}^{2}-A_{1}^{2})
\end{aligned}
\end{equation}

In the case $\delta=0$ the dynamical flow is of type  \emph{relaxational
gradient}~\cite{MontagnePhysica,SMTChile}, that is, there exists a Lyapunov
functional ($\mathcal F$)  that monotonically decreases in time. When $\delta
\neq 0$, however, one cannot find generally such a functional and we say that
the system is  \emph{non-potential}.

\begin{figure}
  \begin{center}
  \epsfig{figure=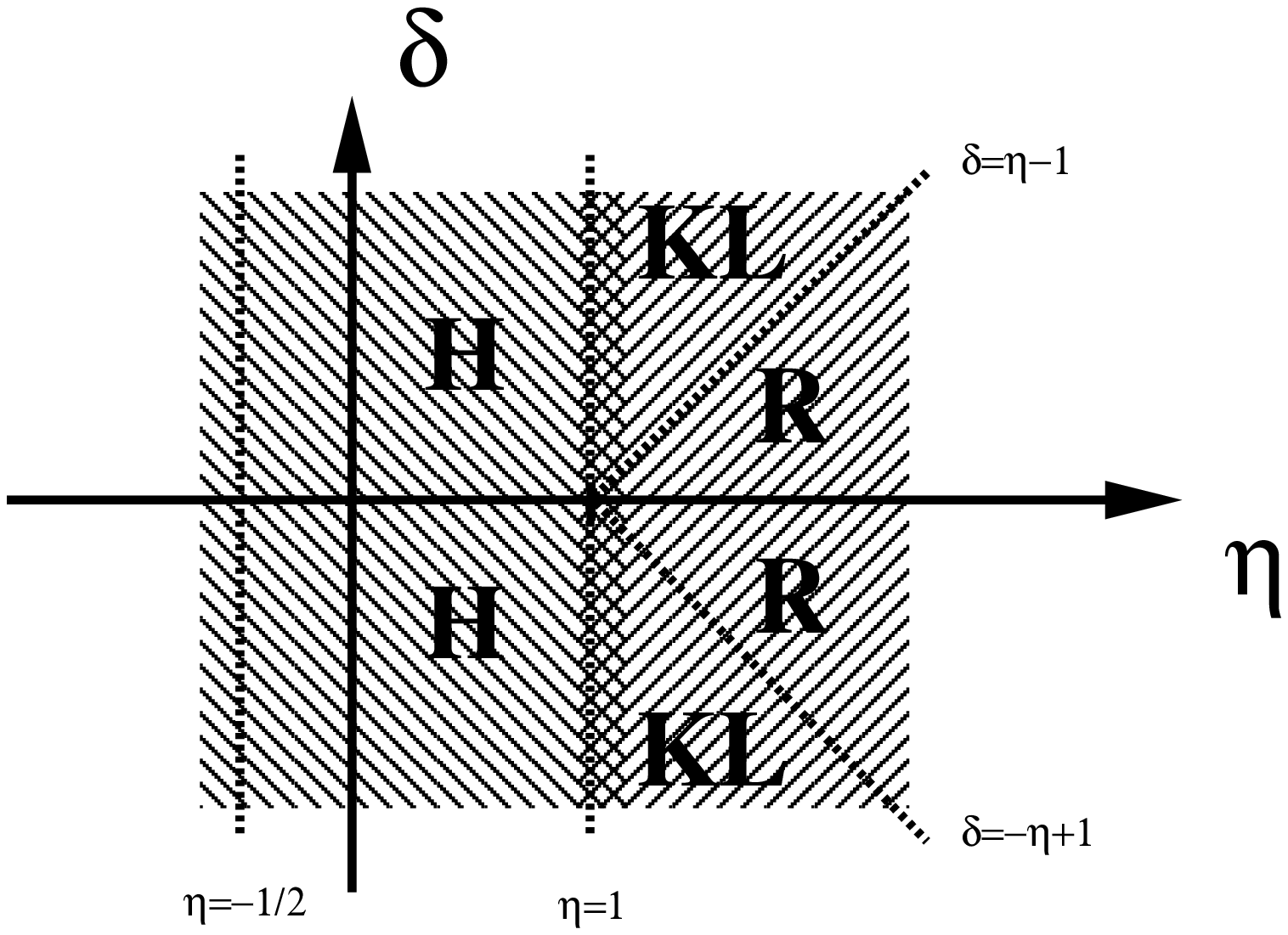,height=9.5cm}
  \vspace{1cm}
  \caption{Linear stability diagram of the homogeneous solutions
  of~(\ref{eq:model}). Inside the region labeled with R, rolls are stable
  whereas in the H region, the stable solution is the hexagon. 
  The KL region corresponds to the K\"uppers--Lortz instability.
  \label{fig:stability}}
  \end{center}
\end{figure}

There exist two kinds of homogeneous solutions which are stable in some region
of the parameter space spanned by $\eta$ and $\delta$. These are three ``roll''
solutions $A_{i}=1,\ A_{j}=0\ (j\neq i),\ i=1,2,3$, and one ``hexagon'' solution
$A_{1}=A_{2}=A_{3}=1/\sqrt{1+2 \eta}$ (this solution requires $\eta > -1/2$).
A linear stability diagram of these solutions is shown in
figure~\ref{fig:stability}. We have focused on the region labeled with the
letter `R' for rotation speeds below the Kuppers-Lortz instability ($|\delta|<
\eta-1$) and where the rolls are the stable solutions. In this region we
expect the formation of domain walls connecting homogeneous stable roll
solutions.

%%%%%%%%%%%%%%%%%%%%%%%%%%%
\section{Front solutions}
\label{Front}
%%%%%%%%%%%%%%%%%%%%%%%%%%%

%%%%%%%%%%%%%%%%%%%%%%%%%%%%
\subsection{Isolated fronts}
%%%%%%%%%%%%%%%%%%%%%%%%%%%%

In the context of the present study, fronts or domain walls are defects that
connect two stable homogeneous solutions. Fronts in one dimension are usually
termed \emph{kinks} and  we will often refer to them in this way.  We focus on
the spatial-dependent stationary solutions of (\ref{eq:model}) with $\delta =
0$. The stable kink solutions are such that one of the three amplitudes, say
$A_{k}$, satisfying the boundary conditions $A_{k}(x\rightarrow\pm\infty)=0$ 
is zero everywhere. In order to study the dynamics of the non-potential kinks,
we first consider the kinks associated with the stationary potential problem
and then we treat the non-potential terms as a perturbation. The two
non-vanishing stationary amplitudes $A_{i}$ and $A_{j}$ are, for $\delta=0$, 
solutions of:
\begin{equation}\label{eq:var}
  \begin{aligned}
    \partial_{x}^{2}A_{i} &= -A_{i}+A_{i}^{3}+\, \eta\, A_{i} A_{j}^{2}	\\
    \partial_{x}^{2}A_{j} &= -A_{j}+A_{j}^{3}+\, \eta\, A_{j} A_{i}^{2}
  \end{aligned}
\end{equation}
with boundary conditions $A_{i}(-\infty)=A_{j}(+\infty)=0$ and
$A_{i}(+\infty)=A_{j}(-\infty)=1$.

The system (\ref{eq:var}) may be considered to represent the two dimensional
motion of a Newtonian particle of unit mass  ($x\rightarrow t$,
$A_{i}\rightarrow X$, $A_{j}\rightarrow Y$) under the action  of a force with 
potential function
$V(X,Y)=\frac{1}{2}(X^{2}+Y^{2})-\frac{1}{4}(X^{4}+Y^{4})-\frac{1}{2}\eta
X^{2}Y^{2}$. This function has two maxima in $m_{0}=\{A_{i}=1,A_{j}=0\}$ and
$m_{1}=\{A_{i}=1,A_{j}=1\}$. It is clear that there exists a unique trajectory
(allowed by the dynamics) along which a particle located in $m_{0}(m_{1})$ can
reach $m_{1}(m_{0})$. The kink profile corresponds to the variation in time
of the particle coordinates $(X(t),Y(t))$ when it moves between the two
maxima~\cite{Chan}.

An explicit analytical solution can be found in two particular
cases~\cite{Malomed}. First, when $0<\eta -1\ll 1$, we have:
\begin{equation}\label{eq:analytical1}
  \begin{aligned}
    A_{i}^{0}(x)&=r(x)\, \left[ 1+\exp \left( 2\sqrt{\eta-1}(x-x_{0}) \right)
    \right]^{-1/2}  \\
    A_{j}^{0}(x)&=r(x)\,e^{x}\, \left[ 1+\exp \left( 2\sqrt{\eta-1}(x-x_{0}) \right)
    \right]^{-1/2}
  \end{aligned}
\end{equation}
with $r(x)=1+(\eta-1)R(x),\ R(x)=O(1)$. Secondly, when $\eta=3$ it is
possible to obtain exact analytical solutions:
\begin{equation}\label{eq:analytical2}
  \begin{aligned}
    A_{i}^{0}(x)&=
    \frac{1}{1+e^{\mp\sqrt{2}(x-x_{0})}}=
    \frac{1}{2} \left[ 1 \pm \tanh \left(
    \frac{x-x_{0}}{\sqrt{2}} \right) \right]
    \\
    A_{j}^{0}(x)&=
    \frac{1}{1+e^{\pm\sqrt{2}(x-x_{0})}}=
    \frac{1}{2} \left[ 1 \mp \tanh \left(
    \frac{x-x_{0}}{\sqrt{2}} \right) \right]
  \end{aligned}
\end{equation}
In both cases $x_0$ is arbitrary but fixed. From these solutions  it is clear
that the spatial scale over which $A_{i}^{0}$ and $A_{j}^{0}$ vary is of order
$1/\sqrt{\eta-1}$.

The three roll solutions are equivalent and they yield the same value for the
Lyapunov functional (\ref{eq:Lyap}). Therefore, we expect kinks not to move in
the potential problem ($\delta=0$). We now ask about the persistence of these
kink solutions when $\delta$ is different from zero. For this we will use
singular perturbation theory. Let us assume $\delta$ to be small, say of order
$\eps$, and look for a solution of (\ref{eq:model}) (with $A_{k}(x)=0$) of the
form:
\begin{equation}\label{eq:trial}
	\begin{aligned}
		A_{i}(x)&=A_{i}^{0} (x-s(t))+\eps\, A_{i}^{1} (x-s(t))+ O(\eps^2)	\\
		A_{j}(x)&=A_{j}^{0} (x-s(t))+\eps\, A_{j}^{1} (x-s(t))+ O(\eps^2)
	\end{aligned}
\end{equation}
where $A_{i}^{0}(x)$ and $A_{j}^{0}(x)$ are solutions of (\ref{eq:var}).
Substituting into (\ref{eq:model}) and matching the terms of the same
order in $\eps$, we find, at order $O(\eps^0)$:
\begin{equation}\label{eq:order0}
  \begin{aligned}
    \partial_{x}^{2}A_{i}^{0}+A_{i}^{0}-(A_{i}^{0})^{3}-
    \,\eta\, A_{i}^{0}(A_{j}^{0})^{2}\, &=0 \\
    \partial_{x}^{2}A_{j}^{0}+A_{j}^{0}-(A_{j}^{0})^{3}-
    \,\eta\, A_{j}^{0}(A_{i}^{0})^{2}\, &=0
  \end{aligned}
\end{equation}
and at order $O(\eps^{1})$:
\begin{equation}\label{eq:order1}
	\mathcal{L} \, \bsy{\alpha}=\bsy{\alpha}'
\end{equation}
where
\begin{gather*}
  {\mathcal L}=\begin{pmatrix}
  \partial_{x}^{2}+1-\eta\,[(A_{i}^{0})^{2}+(A_{j}^{0})^{2}] &
  -2\eta A_{i}^{0}A_{j}^{0}	\\
  -2\eta A_{i}^{0}A_{j}^{0}	& \partial_{x}^{2}+1-
  \eta\,[(A_{i}^{0})^{2}+(A_{j}^{0})^{2}]	
		   \end{pmatrix}
\\
  \bsy{\alpha}= \begin{pmatrix} A_{i}^{1} \\ A_{j}^{1} \end{pmatrix},\
  \bsy{\alpha}'=\begin{pmatrix}
	  \delta{\eps}^{-1}(A_{i}^{0})^{2}A_{j}^{0}-A_{i}^{0}\partial_{t}s \\
	  -\delta{\eps}^{-1}A_{j}^{0}(A_{i}^{0})^{2}-A_{j}^{0}\partial_{t}s
		    \end{pmatrix}
\end{gather*}
The solvability condition for the existence of a solution $(A_{i}^{1}(x),\
A_{j}^{1}(x))$ reads
\begin{equation}\label{eq:sc}
	(\phidag,\ \bsy{\alpha}')=0
\end{equation}
where $(\cdot,\cdot)$ is a scalar product in $L^{2}(\RE)$ defined by
$(f,g)=\int_{-\infty}^{\infty}\!dx f(x)^{*}g(x)$ and $\phidag$ belongs to the
null space of the auto-adjoint operator ${\mathcal L}$. Because of the
translational invariance, ${\mathcal L}$ has a zero eigenvalue so that its
kernel is not empty. The associated eigenvector is generally known to be 
\begin{equation}
	\phidag=
	\begin{pmatrix}
		\partial_{x}A_{i}^{0} \\ \partial_{x}A_{j}^{0}
	\end{pmatrix}
\end{equation}
This is immediately seen taking, for example, the derivative of  
(\ref{eq:order0}) with respect to $x$. Eq. (\ref{eq:sc}) can now be explicitly
evaluated. From this equation , the  solitary  \emph{kink velocity} in the
non-potential case is obtained at leading order:
\begin{equation}\label{eq:vel}
	v(\delta)\equiv\partial_{t}s=\delta \, \frac
	{\int_{-\infty}^{\infty}dx\,
	A_{i}^{0}A_{j}^{0}\, (A_{j}^{0}\,\partial_{x}A_{i}^{0}-A_{i}^{0}\,\partial_{x}A_{j}^{0})}
	{\int_{-\infty}^{\infty}dx
	\left[ (\partial_{x}A_{i}^{0})^{2}+(\partial_{x}A_{j}^{0})^{2} \right]}
	+O(\delta^{2})
\end{equation}

Therefore, in the non-potential case, the kink moves despite connecting states
associated with the same value of the Lyapunov potential of the equilibrium
problem, as already known for other problems~\cite{Coullet1}. For the
particular case of $\eta=3$ for which an analytical result is available for the
kink profile, (eq. (3.3)), an explicit result is obtained for  the solitary
kink velocity, namely $v(\delta)=\delta \sqrt{2}/4$

\begin{figure}
  \begin{center}
  \epsfig{figure=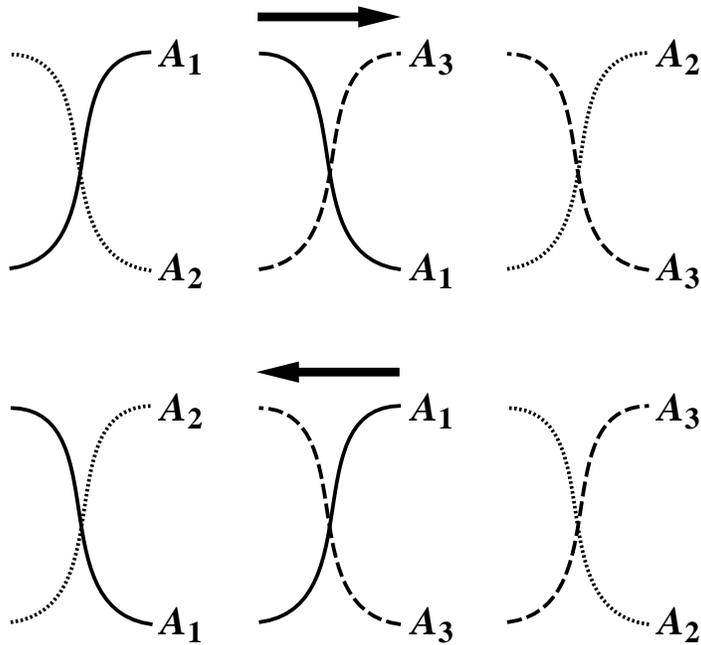,height=9.5cm}
  \caption{Kinds of fronts and their direction of motion for $\delta>0$. The
  remaining amplitude for each kink is understood to be zero across the
  interface. For $\delta<0$ the picture is the same but with the arrows
  interchanged.
  \label{fig:kindfronts}}
  \end{center}
\end{figure}

The expression (\ref{eq:vel}) gives not only the magnitude of the velocity but
also the direction of motion, which is related to the sign of $v$. First, we
note that the velocity is at leading order proportional to $\delta$, so the
direction of the motion depends upon the sign of $\delta$. To illustrate
how (\ref{eq:vel}) determines the direction, let us consider for example a kink
with boundary conditions: $A_{i}(-\infty)=A_{j}(+\infty)=0$ and
$A_{i}(+\infty)=A_{j}(-\infty)=1$; $A_{i}^{0}(x)$ and $A_{j}(x)$ are such that 
$\partial_{x}A_{i}^{0}>0$ and $\partial_{x}A_{j}^{0}<0$. In this case the
numerator of (\ref{eq:vel}) is positive and $v$ has the sign of $\delta$. A
positive  (negative) value  of $v$ corresponds to a kink moving to the right
(left). In figure~\ref{fig:kindfronts} we show a classification of the six
possible types of isolated kinks and their direction of motion. Three of them
move in one direction and the other three in opposite direction.

\begin{figure}
  \begin{center}
  \epsfig{figure=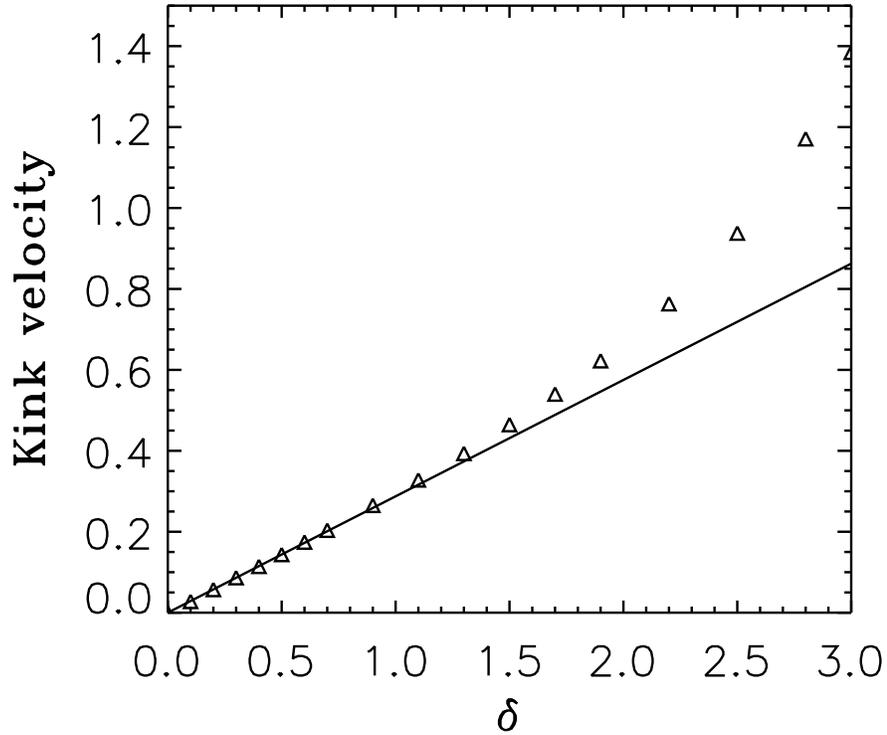, height=9.5cm}
  \vspace{1cm}
  \caption{Solitary kink velocity as a function of the non-potential parameter
  $\delta$
  for $\eta=3.5$. The straight line corresponds to the theoretical perturbative
  approach~(\ref{eq:vel}) whereas the points coming from numerical
  simulation. \label{fig:checkvel}}
  \end{center}
\end{figure}

We have checked numerically the domain of validity of the perturbative result
(\ref{eq:vel}) (see figure~\ref{fig:checkvel}). To check (\ref{eq:vel}) we
either use the analytical result of the  kink profile $A_{i}^{0}$ for $\eta=3$,
or, more generally, the kink profile $A_{i}^{0}$ obtained numerically. For a
value of $\eta=3.5$, we see that the perturbative result to first order in
$\delta$ (\ref{eq:vel}) turns out to be in good agreement with the numerical
results approximately for values of $\delta\lesssim 1.5$. Of course this upper
limit of validity depends on $\eta$ in such way that it gets bigger as $\eta$
is larger. Above this limit the linear relation between $v$ and $\delta$ is no
longer valid and one needs to compute further corrections  in terms of
successive powers of $\delta$.

%%%%%%%%%%%%%%%%%%%%%%%%%%%%%%%%%%%%%%%%%%%
\subsection{Multifront configurations}
\label{multidefect}
%%%%%%%%%%%%%%%%%%%%%%%%%%%%%%%%%%%%%%%%%%%

To study transient dynamics and domain growth we consider random
initial conditions of small amplitude around the unstable solution
$A_{1}=A_{2}=A_{3}=0$. In this situation a multifront pattern emerges rather
than a solitary kink. In order to study dynamical scaling, we are interested in
the late stage  of this dynamics, once well-defined domains  have been formed.

In a potential system governed by a non-conserved scalar order parameter with
short range interactions, as it the case with $\delta=0$, late time dynamics
can be explained in terms of the interaction (and further annihilation) among
adjacent kinks~\cite{Frachebourg,Rutenberg,Nagai,KawasakiOhta}. An isolated
kink is stable. The interacting force between kinks turns out to be
proportional to $\exp (-\alpha d)$~\cite{KawasakiOhta},  where $\alpha$ is some
\emph{positive} constant related to the interface width and other system
parameters, and $d$  is the distance between two adjacent kinks.  This
interaction among kinks leads to a growth law for the characteristic length
that depends on time logarithmically~\cite{Rutenberg,Nagai}. The force  is
attractive and leads to kink annihilation. The process occurs in such way that
the domain that vanishes first is the smallest one. Kink annihilation occurs in
a very small time scale. In fact, the hypothesis of ``instantaneous
annihilation'' has been found to be a good assumption~\cite{Rutenberg}. Kink
annihilation induces domain coarsening leading to a final state with an
homogeneous roll solution filling up the whole system. 

\begin{figure}
  \begin{center}
  \epsfig{figure=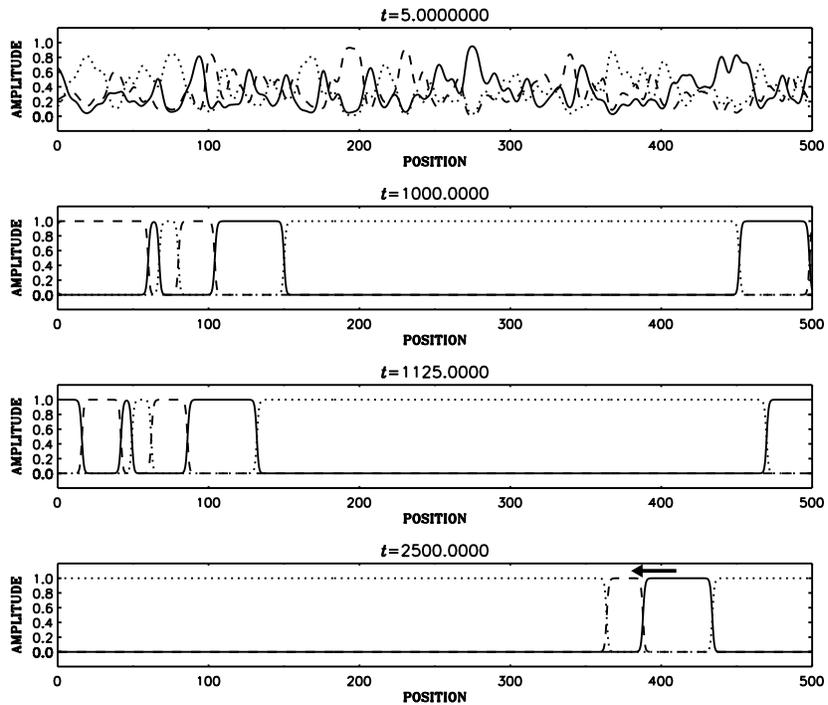, height=9.5cm}
  \vspace{1cm}
  \caption{Snapshots of the temporal evolution of the system. Parameter
  values: $\eta=3.5$, $\delta=0.5$, $L=500$.
  \label{fig:film}}
  \end{center}
\end{figure}

When $\delta$ is different from zero the long stage dynamics should still be
explained in terms of moving fronts which annihilate each other. But now two
very distinct competing physical phenomena come into play. On the one hand,
there is the aforementioned kink interaction. On the other hand, we have the
kink motion driven by non-potential effects. In this case, we do not expect the
growth law to be logarithmic, at least in the regime where the non-potential
effects (the strength of which is measured by $\delta$) are important. In
figure (\ref{fig:film}) we show some snapshots corresponding to a typical run
of the temporal evolution of the system (we use periodic boundary conditions).
The first snapshot corresponds to an early stage during which domains are
forming. Once formed, kinks move in such a  way that annihilations of
contrapropagating adjacent kinks leads to coarsening. Eventually, as
corresponding to the last snapshot, the system may be in a state with a group
of kinks moving all in the same direction. These will interact among them (with
a interaction force that varies logarithmically with the interkink distance)
until extinction.

We have performed a perturbation analysis of domain growth in the simplest
example of a single domain bounded by two moving domain walls. A differential
equation for the  domain size $s(t)$ can be obtained in the general case (see
appendix). For $\eta=3$ it adopts the simple  form:
\begin{equation}\label{eq:size}
	\partial_{t}s(t)=2 v(\delta)-24 \sqrt{2} e^{-\sqrt{2}s(t)}
\end{equation}
where $v(\delta)$ is the solitary kink velocity. This expression is obtained in
the ``dilute-defect gas approximation'', that is, when the width of the fronts
is much smaller than the distance between them. The first term in the right
hand of~(\ref{eq:size}) can be either negative or positive and represents the
contribution to the variation of the domain size owing to non-potential
effects. The second term is related to the interacting force between  the kinks
and it is always negative (attractive force) so that it tends to shrink the
domain. If both terms are negative, the kinks will annihilate each other.
Otherwise, when the first term is positive, the two effects act in opposite
directions. In fact, given an initial size of the domain $s_{0}$, it is
possible to find a value $\delta=\delta_{c}$ for which the domain neither
shrinks nor grows; in this case, the initial domain would not evolve in time
being a stationary solution. For values $\delta > \delta_{c}$ the domain would
get wider whereas for $\delta < \delta_{c}$ it would shrink. Note that in the
$\delta=0$ case (potential regime), the isolated domain always collapses, but
this can be stopped with a suitable strength of the non-potential terms.
Furthermore, given a fixed value of $\delta$, if $s_{0}$ is large enough, the
dominant term responsible of the kink motion is the one associated with
$v(\delta)$. In this case the fronts move at a constant velocity leading to a
variation of the domain size linear with time. On the other hand, if $s_{0}$ is
small enough, kink interaction will be the dominant effect and the single
domain size will collapse logarithmically with time. This picture  of the size
dynamics of a single domain also explains basically what happens when more
domains (and a non-vanishing third amplitude) coexist. It gives a useful
understanding of the characteristic growth laws obtained from a statistical
analysis in the next section. 

%%%%%%%%%%%%%%%%%%%%%%%%%%%%%%%%%%%%%%%%
\section{Domain growth and scaling}
\label{Growth}
%%%%%%%%%%%%%%%%%%%%%%%%%%%%%%%%%%%%%%%%

In this section we focus on the scaling properties of the system
(\ref{eq:model}) in a late stage of the dynamics, namely when well-defined
domains have formed. The scaling hypothesis states that there exists a single
characteristic length scale $R(t)$ such that the domain structure is, in a
statistical sense, independent of time when lengths are scaled by $R(t)$. We
will refer to the time dependence of the scale length as the \emph{growth law}
of the system. It has been found that the scaling hypothesis holds in a great
variety of potential systems. The system under study here gives us the
opportunity of  answering the question of whether a non-potential dynamics
satisfies dynamical scaling.

Two magnitudes frequently used to study domain growth and scaling properties
for a scalar field   $\Psi({\mathbf x},t)$ (for instance, one of the three
amplitudes in equations (2.2)) are the equal time correlation function
\begin{equation}
  C({\mathbf r},t)=\left\langle 
  \sum_{{\mathbf x}}\Psi({\mathbf x}+{\mathbf r},t) \Psi({\mathbf x},t) 
  \right\rangle_{\mathrm i.c.}
\end{equation}
and its Fourier transform, the equal time structure factor
\begin{equation}
  S({\mathbf k},t)=\left\langle 
  \sum_{{\mathbf k}}\hat{\Psi}({\mathbf k},t) \hat{\Psi}(-{\mathbf k},t) 
  \right\rangle_{\mathrm i.c.}
\end{equation}
where the angular brackets indicate an average over initial conditions
(``runs''). If a single characteristic length exists, according to the scaling
hypothesis, the pair correlation function and the structure factor must have
the following scaling forms in a $d$-dimensional system:\\ 
\begin{align}\label{eq:scalinghypothesis}
  C({\mathbf r},t)&=f(r/R(t)) \\
  S({\mathbf k},t)&=R(t)^{d}\hat{f}(kR(t))
\end{align}
The function $f$ is called the \emph{scaling function}. To check numerically
the validity of the  previous scaling laws, we have integrated the system of
equations~(\ref{eq:model})  using a finite difference method for both, the
spatial and temporal derivatives. In the simulations we have taken a constant
value for $\eta$, namely $\eta=3.5$, and we have varied $\delta$ from
$\delta=0$ (potential case) to $\delta=0.1$ (a value below the KL instability
threshold).  We  have used periodic boundary conditions and have averaged our
results over 100--500 runs and used system sizes ranging from $L=250$ to
$L=1000$. To study domain structure, we consider the correlation function of
one of the three amplitudes. We use the correlation function better than the
structure factor because of the large fluctuations of the structure factor at
small wave numbers. A typical length scale associated with the average domain
size can be defined in several ways. Specifically, we have determined it by
computing the value of $r$ for  which $C(r,t)$ is half its value at the origin
at time $t$, that is $C(R(t),t)={\scriptstyle \frac{1}{2}}C(0,t)$. The
calculation has been performed by fitting the four points of $C(r,t)$ closest
to $C(0,t)$ to a cubic polynomial. Another typical length, $R_{1}(t)$ can be
evaluated  directly as the system size divided by the number of kinks. We have
verified that the quotient  $R_{1}(t)/R(t)$ remains nearly constant, as
expected, when a single characteristic  length dominates the problem.

%%%%%%%%%%%%%%%%%%%%%%%%
\subsection{Growth Law}
%%%%%%%%%%%%%%%%%%%%%%%%

\begin{figure}
  \begin{center}
  \epsfig{figure=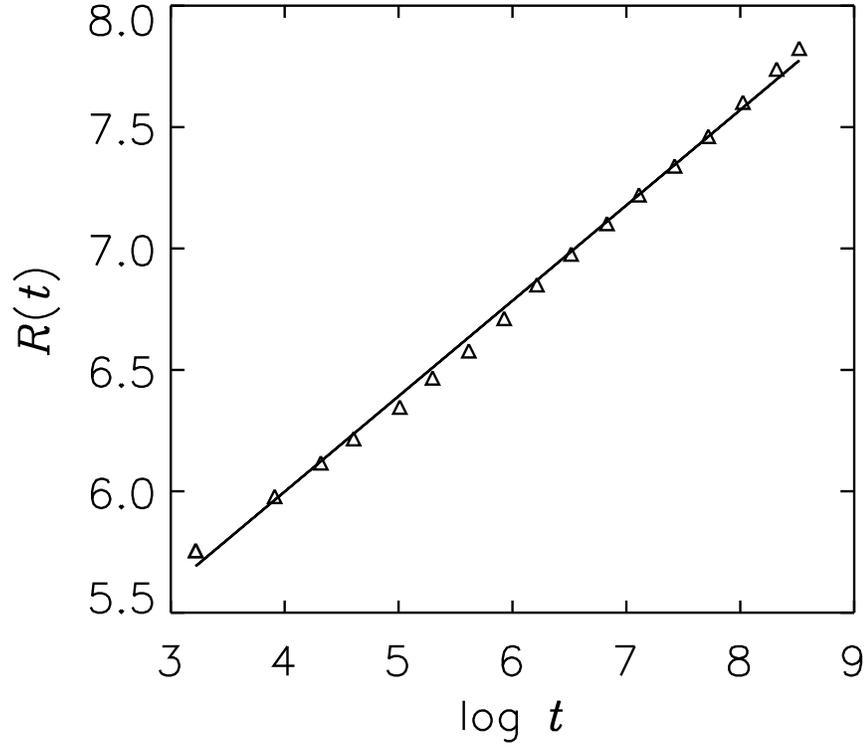, height=9.5cm}
  \vspace{1cm}
  \caption{Time evolution of the characteristic domain size for the potential
  case $\delta=0$ and $L=1000$.  The straight line is a
  linear regression fit of points obtained numerically. \label{fig:GLvar}}
  \end{center}
\end{figure}

We consider first the potential case $\delta=0$: In figure~\ref{fig:GLvar} we
show that the domain size follows the expected logarithmic behavior. The
attractive interaction among the kinks leads to a very long transient before
the system reaches its final state which corresponds to one roll solution
filling up the whole system.

\begin{figure}
  \begin{center}
  \epsfig{figure=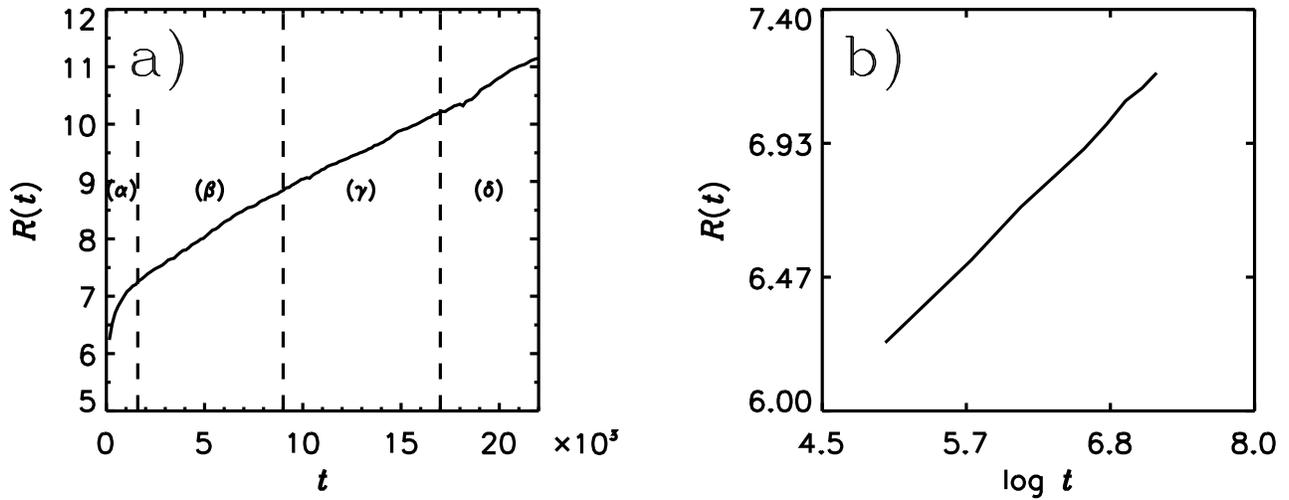, width=\textwidth}
  \vspace{1cm}
  \caption{(a) Time evolution of the characteristic domain size for $\delta=0.001$
  and $L=1000$. The initial logarithmic growth law (region ($\alpha$)) becomes
  linear (region ($\gamma$)) after a crossover (region ($\beta$)). Region
  ($\delta$) is related to finite size effects. (b) Zoom of region ($\alpha$) in
  the left plot with logarithmic time scale. 
  \label{fig:GLdeltasmall}}
  \end{center}
\end{figure}

In the non-potential case the domain size $R(t)$ is shown in
figure~\ref{fig:GLdeltasmall} for a system of size $500$ and $\delta=10^{-3}$.
For the earliest times, when the kinks are very close to each other, and
according to the  discussion in section~\ref{multidefect} we expect the
interaction terms to be the dominant ones (as long as  $\delta$ is small
enough). This leads to a logarithmic growth of $R(t)$ as observed in region
($\alpha$) of  figure~\ref{fig:GLdeltasmall}. Due to coarsening the
characteristic domain size becomes larger and the domain wall interaction
becomes weaker as the time increases.  For longer times the non-potential
effects dominate with respect to wall interaction. In this  regime we can
consider each domain wall to move at a constant velocity. This gives rise to a
linear behavior of  $R(t)$ with time (region ($\gamma$)). Between regions
($\alpha$) and ($\gamma$) there exists a crossover (region ($\beta$)) for which  the
weights of both effects (interaction and non-potential) in driving the domain
wall motion are of the same  order. Finally, at very late times finite size
effects come into play (region ($\delta$)): the domain size saturates  to a
constant value and the number of domain walls is too small to make good
statistics.

In the regime for which the interaction effects are the dominant ones, we can
give a simple explanation of  the linear growth law observed for $R(t)$.
Statistically speaking, there will be the same number of kinks  moving to the
right and to the left. As a matter of fact, in an appropriate reference frame,
the system can be seen as composed of motionless kinks (type 1) and kinks
moving at a velocity of  $2v$ in one fixed direction (type 2) . If we call
$N_{1}(t)$ and $N_{2}(t)$ the \emph{average} number of kinks of both types at
time $t$, the number of kinks of, say type 1, at time $t+dt$ will be:
\begin{equation}\label{eq:n1n2}
  N_{1}(t+dt)=N_{1}(t)-N_{1}(t)\, v \, dt\times\frac{N_{2}(t)}{L}
\end{equation}
where the second term in the right hand side represents the number of kinks
disappeared in $dt$ by annihilation and $L$ stands for the system size. The
important point is that \mbox{$N_{1}(t)=N_{2}(t)=N(t)$} (remember we are
dealing with averaged quantities), so that (\ref{eq:n1n2}) transforms into:
\begin{equation}
  \frac{dN(t)}{dt}=1-\frac{v}{L}N(t)^{2}
\end{equation}
The integration of the previous equation gives
$N(t)=[(v/L)t+N_{0}^{-1}]^{-1}\sim t^{-1}$, so that the average inter-kink
distance $R(t)\sim N(t)^{-1}\sim t$ is linear with time.

\begin{figure}
  \begin{center}
  \epsfig{figure=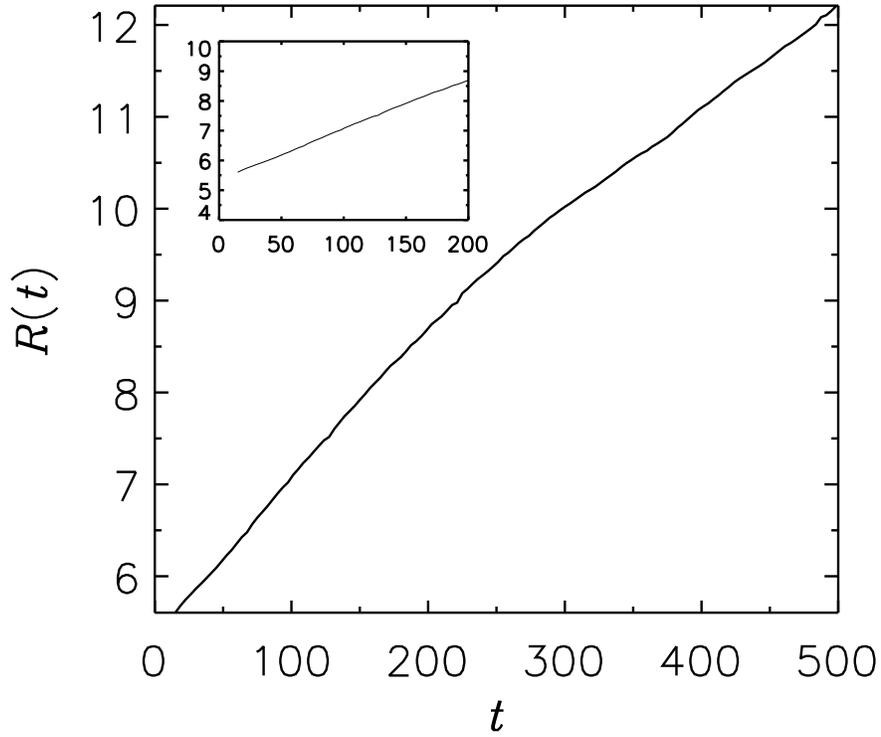, height=9.5cm}
  \vspace{1cm}
  \caption{Time evolution of the characteristic domain size for $\delta=0.1$
  and $L=1000$.
  \label{fig:GLdeltabig}}
  \end{center}
\end{figure}

When $\delta$ is large enough the initial kink annihilation is so fast that the
regions ($\alpha$) and ($\beta$) in the $R(t)$ plot can hardly be observed in the numerical
integration. In this case of  large non-potential effects, a linear growth law
is observed for the shortest times and it continues until finite size effects
show up clearly (see figure~\ref{fig:GLdeltabig}). Finite size effects are seen
to occur for $t>200$ for $\delta=0.1$. The bigger $\delta$, the
sooner finite size effects appear.

At long times the system will consist of an homogeneous roll state or a group
of kinks moving either to the right or to the left \cite{foot1}. Note that the
periodic boundary conditions impose constraints about the number of such moving
kinks. To be precise, the number of kinks moving in a fixed direction must be
multiple of three. We can form a subgroup of three kinks moving in the same
direction by joining those appearing in each row of
figure~(\ref{fig:kindfronts}). The moving kinks will carry on interacting among
them (logarithmically) until eventually they all will disappear. In this
situation we expect the growth law to be logarithmic with time but one of such
groups is composed typically of three, six or rarely nine kinks, a number too
small as to make good statistics.

%%%%%%%%%%%%%%%%%%%%%%%%%%%%%
\subsection{Scaling function}
%%%%%%%%%%%%%%%%%%%%%%%%%%%%%

\begin{figure}
  \begin{center}
  \epsfig{figure=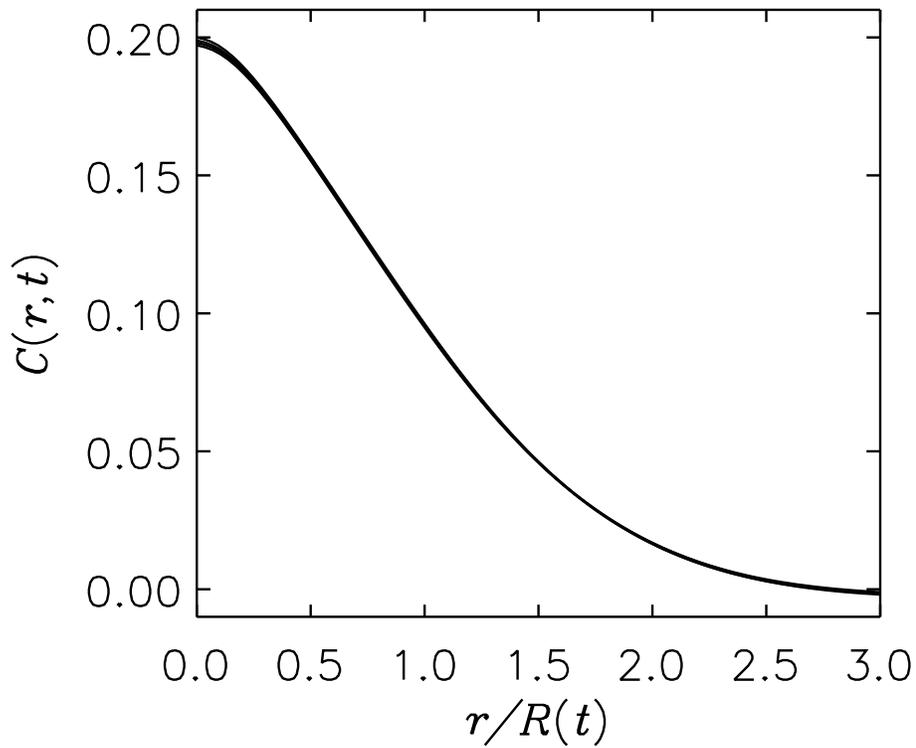, height=9.5cm}
  \vspace{1cm}
  \caption{Scaling function for the potential case $\delta=0$. The plot has been
  made by
  over plotting $C(r,t_{i})\ \text{vs.}\ r/R(t_{i})$ for several
  times from $t=200$ up to $t=5000$.
  \label{fig:CFvar}}
  \end{center}
\end{figure}

\begin{figure}
  \begin{center}
  \epsfig{figure=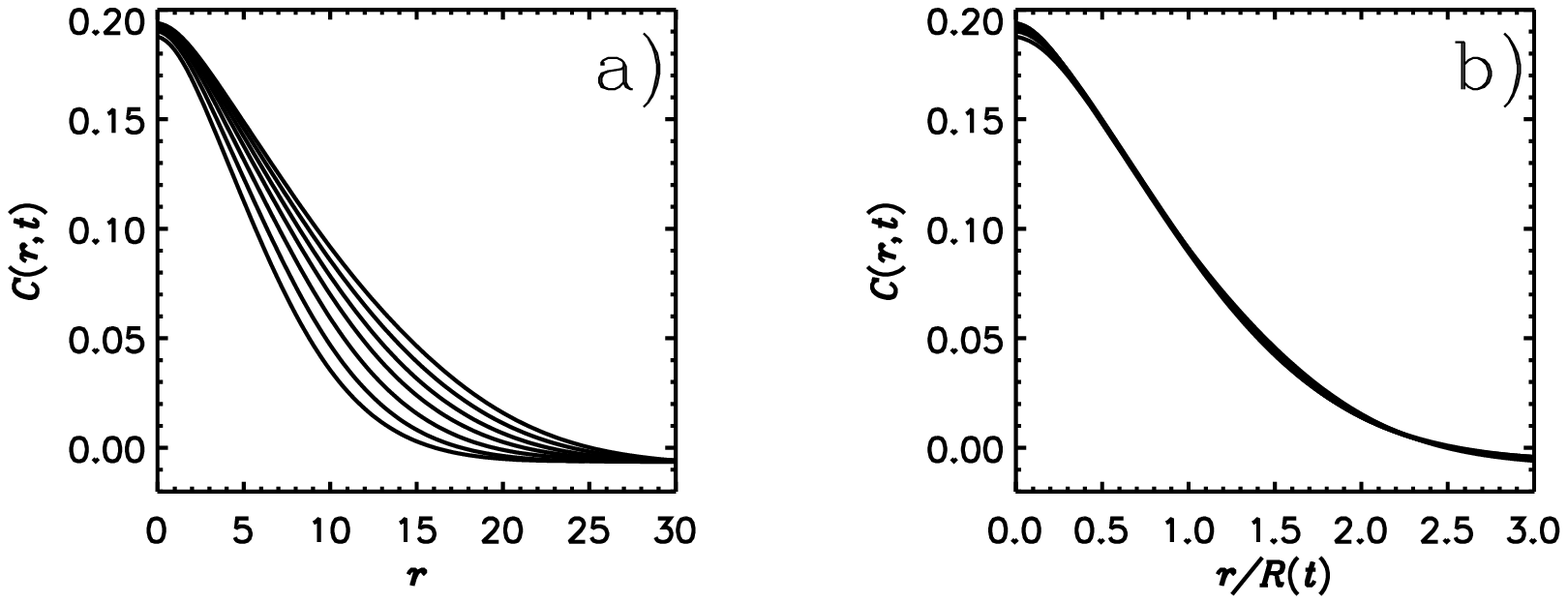, width=\textwidth}
  \vspace{1cm}
  \caption{(a) Equal time correlation function versus the non-scaled length for
  $\delta=0.001$, $L=1000$ and several different times from t=150 to
  t=15000. (b) Equal time correlation function versus the scaled length. The
  system parameters and the times for each curve are the same as in figure (a). 
  \label{fig:CFdeltasmall}}
  \end{center}
\end{figure}

\begin{figure}
  \begin{center}
  \epsfig{figure=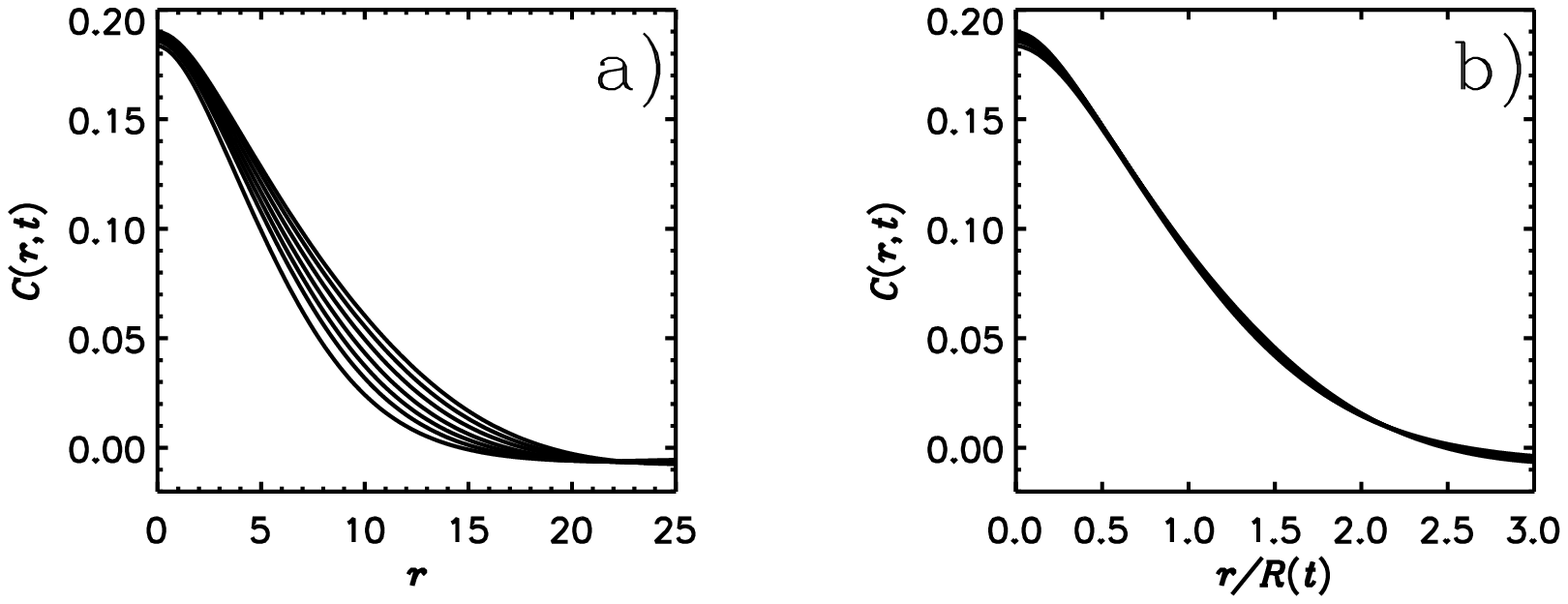,width=\textwidth}
  \vspace{1cm}
  \caption{(a) Equal time correlation function versus the non-scaled length for
  $\delta=0.1,\ L=1000$ and several different times from t=15 to t=150. (b) 
  Equal time correlation function versus the scaled length. The
  system parameters and the times for each curve are the same as in figure (a).
  \label{fig:CFdeltabig}}
  \end{center}
\end{figure}

We now address the question of the validity of the dynamical scaling hypothesis
(\ref{eq:scalinghypothesis}).  For this purpose, we have plotted the equal time
correlation function $C(r,t)$ versus the scaled length $r/R(t)$ for several
times. Figure~\ref{fig:CFvar} shows the scaling function in the potential case.
In figures \ref{fig:CFdeltasmall} and \ref{fig:CFdeltabig} we show the 
correlation functions for several times before and after scaling the system
length. Our results show that the correlation functions follow a single profile
when the length is scaled with the characteristic domain sizes obtained above. 
We therefore conclude that a scaling description of the system is also valid as
in the potential case, but  now with a non-potential dynamics. The upper limit
of the time interval during which there is scaling is determined by the
appearance of finite size effects. The range of values of  $\delta$ for which
there is scaling in a quite large time interval  is rather small. For values
of $\delta$ even of a few tenths, the finite size effects show up for very
short times. Moreover  the fluctuations in the scaling function grow as
$\delta$ increases. For these reasons, we have not been able to obtain a
conclusive comparison among scaling functions for different values of 
$\delta$, although their shapes appear to be rather
insensitive to the value of $\delta$.  

%%%%%%%%%%%%%%%%%%%%%%%
\section{Conclusions}
\label{theend}
%%%%%%%%%%%%%%%%%%%%%%

We have studied domain growth and dynamical scaling in a non-potential 
coarsening process in one dimension. The model considered is motivated by the
phenomenon of  Rayleigh-B\'enard convection in a rotating cell.  We have
focused on the region below the K\"uppers-Lortz instability point, where the
dynamics is  still non-potential and  the system shows coarsening. A solitary
kink moves at a constant velocity due to the non-potential dynamics. When there
are several kinks present in the system, these move due to both, domain wall
interaction and  non-potential effects. In any case the dynamics is governed by
motion of interfaces. This motion is such that  kinks moving in opposite
directions annihilate each other. As a consequence of kink annihilation the
average  domain size grows in time and the system coarsens. When $\delta=0$
(potential case) we have shown that, in accordance with general results, the
growth law is logarithmic with time and that a scaling description of the
system dynamics is possible.  When $\delta$ becomes different from zero we have
found that the scaling hypothesis still holds, as in the  potential case, but
with a different growth law that reflects the non-potential dynamics of the
system. For the  shortest times, the kink interaction (the only effect present
in the potential case) is the dominant effect and  gives rise to a logarithmic
growth law with time. For longer times the average inter-kink  distance is
large enough to make the interaction effects negligible in driving  kink
motion.  Therefore each kink moves nearly independently of the others as if it
were isolated. In this situation  each  kink  moves with a constant velocity
leading to a linear growth law. For larger values of $\delta$ the logarithmic 
region is not  observed because of the fast annihilation of the domains during
the very early times. The two dimensional version of this problem is currently
under study  \cite{2d} and it exhibits rather different dynamical behavior
grossly dominated by vertices where three domain walls meet and which have no
parallel in one dimensional systems.

%%%%%%%%%%%%%%%%%%%
\section*{APPENDIX}
\appendix
\renewcommand{\theequation}{A.\arabic{equation}}
%%%%%%%%%%%%%%%%%%%

We consider an isolated domain bounded by two domain walls associated with
amplitudes $A_{1}$ and $A_{2}$, while $A_{3}=0$. When the domain size is much
greater than the interface width (``dilute-defect gas approximation''), a
reasonable \emph{ansatz} for this solution is \\
\begin{equation}\label{eq:ansatz}
 \begin{aligned} 
   A_{1}(x,t) & = a(x-r(t))+b(x-d+r(t))+w_{1}(x,t)	\\ 
   A_{2}(x,t) & = b(x-r(t))+a(x-d+r(t))-1+w_{2}(x,t)
 \end{aligned} 
\end{equation}
where $r(t)$ measures the displacement of the kinks, $d$ is the initial domain
size (so that the domain size at time $t$ is $d-2r(t)$), $\partial_{t}r$ and
$w_{i}  \ (i=1,2)$ are assumed to be small corrections of order $\delta$ and
$\partial_{t}w_{i}$ to be negligible with respect to $w_{i}$. To simplify
notation, we use: $f \equiv f(x-r(t)),\ f_{d} \equiv f(x-d+r(t))$. The moving
fronts $a$ and $b$ satisfy the boundary conditions
$a(\infty)=b(-\infty)=0, \ a(-\infty)=b(\infty)=1$ and they are solutions of
the system (\ref{eq:model}) (with one of the amplitudes equal to zero) so that
the following equations hold:
\begin{equation}
  {\mathcal M}_{+}(a,b)={\mathcal M}_{+}(b_{d},a_{d})=
  {\mathcal M}_{-}(b,a)={\mathcal M}_{-}(a_{d},b_{d})=0
\end{equation}
where the action of the operators ${\mathcal M}_{\pm}(\cdot,\cdot)$
is given by:
\begin{equation}
  {\mathcal M}_{\pm}(f,g)=\partial_{x}^{2}\pm v(\delta)\partial_{x}+f-f^{3}
  -(\eta\pm\delta)fg^{2}
\end{equation}
The parameter $v(\delta)$ is the front velocity as given by eq.~(\ref{eq:vel}). 

Introducing the ansatz~(\ref{eq:ansatz}) into (\ref{eq:model}) we obtain, at
leading order, a linear system of equations for $w_{1}$ and $w_{2}$: 
\begin{gather}
  {\mathcal L}\bsy{\phi}=\bsy{\phi}' \label{eq:linw} \\
  {\mathcal L}=
  \begin{pmatrix}
    \partial_{x}^{2}+1-3(a+b_{d})^{2}-\eta(b+a_{d}-1)^{2} & 
    -2 \eta (a+b_{d})(a_{d}+b-1)	\\
    -2 \eta (a+b_{d})(a_{d}+b-1) &
    \partial_{x}^{2}+1-3(a_{d}+b-1)^{2}-\eta(a+b_{d})^{2}
  \end{pmatrix}  \notag \\ 
  \bsy{\phi}=
    \begin{pmatrix}
  	     w_{1}  \\
  	     w_{2}
  \end{pmatrix}, \
  \bsy{\phi}'=
  \begin{pmatrix}
    (\partial_{x}a+\partial_{x}b_{d})v(\delta)+(\partial_{x}b_{d}-
    \partial_{x}a)\partial_{t}r+K_{1}\delta+K_{2}\eta+K_{3}  \\
    (\partial_{x}b+\partial_{x}a_{d})v(\delta)+(\partial_{x}a_{d}-
    \partial_{x}b)\partial_{t}r+K'_{1}\delta+K'_{2}\eta+K'_{3}
  \end{pmatrix}	\notag
\end{gather}
where the functions $K_{i}(x,t)$ and $K'_{i}(x,t) \ (i=1,2,3)$ are given by:
\begin{align*}
  K_{1}&=a(a_{d}-1)^{2}+2b(a+b_{d})(a_{d}-1)+b_{d}(1-2a_{d}+b^{2})  \\
  K_{2}&=a(a_{d}-1)^{2}+2(a_{d}-1)(ab+a_{d}b_{d}+bb_{d})+b_{d}(1+b^{2})  \\
  K_{3}&=-2\,\partial_{x}^{2}b_{d}+b_{d}(3a^{2}+3ab_{d}+2b_{d}^{2}-2) \\
  K'_{1}&=-a(a+2b_{d})(a_{d}-1)+b_{d}(b_{d}-2ab-bb_{d})  \\
  K'_{2}&=a(a+2b_{d})(a_{d}-1)+b_{d}(2ab+bb_{d}+2a_{d}b_{d}-b_{d})  \\
  K'_{3}&=-2\,\partial_{x}^{2}a_{d}+3b^{2}(a_{d}-1)+3b(a_{d}-1)^{2}+2a_{d}^{3}
         -3a_{d}^{2}+a_{d}
\end{align*}
The solvability condition for the existence of a solution
$(w_{1}(x,t),w_{2}(x,t))$ for (\ref{eq:linw}) reads
\begin{equation}\label{eq:sc2}
  (\bsy{\Psi}^{\dag},\ \bsy{\phi}')=0
\end{equation}
where $\bsy{\Psi}^{\dag}$ belongs to the kernel of the auto-adjoint linear
differential operator $\mathcal{L}$. We will show below that
$\bsy{\Psi}^{\dag}$ is  approximately given by $(\partial_{x}a,\
\partial_{x}b)^{T}$ (here $T$ denotes the transposed vector), where
$a=a(x-r(t))$ and $b=b(x-r(t))$ are the domain wall profiles around $x=r(t)$.

The first component of the vector ${\mathcal L}\bsy{\Psi}^{\dag}$
is given by:
\begin{equation}\label{eq:comp1}
  \begin{split}
    ({\mathcal L}\bsy{\Psi}^{\dag})_{1} &=  {\mathcal
    L}_{11}\partial_{x}a+{\mathcal L}_{12}\partial_{x}b \\ 
    &=\partial_{x}^{3}a+\partial_{x}a-3(a+b_{d})^{2}\partial_{x}a-\eta
    (a_{d}+b-1)^{2}\partial_{x}a-2\eta (a+b_{d})(a_{d}+b-1)\partial_{x}b
  \end{split}
\end{equation} 
As long as that the width of the interfaces is much smaller than the domain
size (for all times $t$), we can make the following approximations: $ab_{d}
\approx 0,\ aa_{d} \approx a,\ bb_{d} \approx b_{d}$. Moreover, this assumption
implies that the product of the derivative with respect to $x$ of an amplitude
solution centered on $x=x_{0}$ multiplied by another amplitude shifted a length
of order of the domain size, will be a function which will take values
different from zero only in a small region around $x=x_{0}$. By using the 
approximations
\begin{equation}
\begin{gathered}
  (a+b_{d})^{2}\partial_{x}a \approx a^{2}\partial_{x}a  \\
  (a_{d}+b-1)^{2}\partial_{x}a \approx b^{2}\partial_{x}a \\
  (a+b_{b})(a_{d}+b-1)\partial_{x}b \approx ab\partial_{x}b
\end{gathered}
\end{equation}
we find:
\begin{gather}
  ({\mathcal L}\bsy{\Psi}^{\dag})_{1}=\partial_{x}\left[\partial_{x}^{2}a+a-a^{3}-\,\eta\, b^{2}a\right]
\end{gather}
Taking the derivative of  (\ref{eq:model}) with respect to $x$  we find that
the right hand side of (A.8) is equal to zero when the amplitude solutions
\mbox{$a=a(x-r(t))$} and $b=b(x-r(t))$ are replaced by its form for $\delta=0$.
Hence, we conclude that  $({\mathcal L}\bsy{\Psi}^{\dag})_{1}=O(\delta)$.
Likewise, we can prove that  $({\mathcal L}\bsy{\Psi}^{\dag})_{2}=O(\delta)$.
Therefore,  at lowest order in $\delta$, $(\partial_{x}a,\ \partial_{x}b)^{T}$
belongs to the kernel of the operator  $\mathcal L$. 

%Another element of this kernel However, the vector
%$(\partial_{x}a_{d},\ \partial_{x}b_{d})^{T}$ is also valid, but it can be
%shown that the results obtained with both of them are the same so we can
%use either. 

Now we can calculate the evolution of the domain size $s(t)=d-2r(t)$ through
the solvability condition (\ref{eq:sc2}). We obtain:
\begin{equation}\label{eq:domsize}
   \partial_{t}s \cong \pm 2v(\delta)+\frac
   {
   \int_{-\infty}^{\infty} dx (h_{a}\, \partial_{x}a+h_{b}\, \partial_{x}b)
   }
   {
   \int_{-\infty}^{\infty} dx [(\partial_{x}a)^{2}+(\partial_{x}b)^{2}]
   }
\end{equation}
where the coefficients $h_{a}$ and $h_{b}$ depend upon the amplitude solutions
$a$ and $b$ and the non-potential parameter $\delta$. The first term of the
right hand side of (\ref{eq:domsize}) represents the rate of change of the
domain size due to non-potential effects which cause the kinks to move at a
constant velocity $v(\delta)$. The second term is related to kink interaction. In
the case $\eta=3$ we can compute explicitly all the coefficients  involved in
(\ref{eq:domsize}) taking advantage of the analytical kink profiles at lowest
order in $\delta$  (equation (\ref{eq:analytical2})). Making an expansion in
powers of $e^{-\sqrt{2}s(t)}$, retaining only the leading terms, and provided
that $\delta$ is a small parameter,  we obtain:
\begin{gather}
  \partial_{t}s=\pm \frac{\delta}{\sqrt{2}}-24\sqrt{2}e^{-\sqrt{2}s(t)}
\end{gather}
which is eq.~(\ref{eq:size}).

%%%%%%%%%%%%%%%%%%%%%%%%%%%%%%%%%%%%%%%%%%%%%%%%%%%%%%%%%%%%%%%%%%%%%%%%%%%%%%
%%%%%%%%%%%%%%%%%%%%%%%%%%%%%%%%%%%%%%%%%%%%%%%%%%%%%%%%%%%%%%%%%%%%%%%%%%%%%%

\noindent ACKNOWLEDGMENTS:
Financial support from DGYCIT (Spain) Projects PB94-1167 and PB94-1172 is
acknowledged.

%%%%%%%%%%%%%%%%%%%%%%%%%%%%%%%FIGURES%%%%%%%%%%%%%%%%%%%%%%%%%%%%%%%%%%%

%%%%%%%%%
%FIG1.PS%
%%%%%%%%%
%%%%%%%%%%
%%FIG2.PS%
%%%%%%%%%%
%%%%%%%%%%
%%FIG3.PS%
%%%%%%%%%%
%%%%%%%%%%
%%FIG4.PS%
%%%%%%%%%%
%%%%%%%%%%
%%FIG2.PS%
%%%%%%%%%%
%%%%%%%%%%
%%FIG6.PS%
%%%%%%%%%%
%%%%%%%%%%
%%FIG7.PS%
%%%%%%%%%%
%%%%%%%%%%
%%FIG8.PS%
%%%%%%%%%%
%%%%%%%%%%
%%FIG9.PS%
%%%%%%%%%%
%%%%%%%%%%
%%FIG10.PS%
%%%%%%%%%%
%
%
%%%%%%%%%%%%%%%%%%%%%%%%%%%%%%%


\begin{thebibliography}{10}
\bibitem{foot1} We have checked on our numerical simulations, that on average,
half the runs lead to a final state corresponding to a  group of kinks moving to 
the right and half of them to kinks
moving to the left.
\end{thebibliography}
\end{document}